\documentstyle[11pt]{article}
 
\def\pa{\partial}
\def\k{\kappa} 
\def\g{\gamma} \def\G{\Gamma}
\def\a{\alpha} 
\def\b{\beta} 
\def\d{\delta} 
\def\e{\epsilon} 
 
\def\k{\kappa}
\def\l{\lambda} \def\L{\Lambda}
\def\m{\mu} 
\def\n{\nu}

\def\s{\sigma}

\def\o{\omega} 
\def\be{\begin{equation}}
\def\ee{\end{equation}}
\title{No Cosmological $D$=11 Supergravity}
\author{K. Bautier$^{1}$, S. Deser$^2$, M.
Henneaux$^{1,3}$ and D. Seminara$^2$}

\begin{document}

\maketitle

\begin{center}
{\it $^1$Facult\'e des Sciences, Universit\'e Libre de Bruxelles,
Campus Plaine, \\ C.P. 231, B-1050, Bruxelles, Belgium\\
$^2$ Physics Department, Brandeis University, Waltham, 
MA 02254, USA\\
$^3$Centro de Estudios
Cient\'{\i}ficos de Santiago, Casilla 16443, Santiago, Chile}
\end{center}

\begin{abstract}
We show, in two complementary ways, that $D$=11 
supergravity---in contrast to all its lower 
dimensional versions---forbids a cosmological
extension.  First, we linearize the putative model 
about an Anti de Sitter background and show that
it cannot even support a ``global" supersymmetry
invariance; hence there is no Noether construction that can lead to 
a local supersymmetry. This is true with the usual 
$4-$form field as well as for  a
``dual", 7-form, starting point.  Second, a cohomology
argument, starting from the original full nonlinear theory, 
establishes the absence of deformations  involving
spin 3/2 mass and cosmological terms.  In both approaches, 
it is the form field that is responsible for the obstruction.
``Dualizing'' the cosmological constant to an 11-form field
also fails.
\end{abstract}

%\vspace{.1in}
\vfill

\begin{flushright}
ULB-TH-97/07\\
BRX-TH-411\\
\end{flushright}
\vfill

The recent revival of $D=11$ supergravity \cite{001}
is connected to its role as a sector of
$M$-theory  unification.
Of the many special properties of $D$=11 
supergravity, one  of the most striking is that it {is} 
unique and seems to forbid a cosmological term extension,
which
{\it is} allowed in all lower ($D\leq 10$) dimensions.
In view of the importance of this question to lower-$D$ structures
in the duality context, we propose to establish this obstruction
in a concrete physical way\footnote{ To our knowledge,  
there have been two
previous approaches to  this result. One \cite{Nahm}
consists
in a classification of all graded algebras and
consideration of their highest spin representations.
Although we have not found an explicit exclusion of 
the cosmological extension
in this literature, it is undoubtly implied
there under similar assumptions.
The second 
\cite{sagnotti} considers the properties
of a putative ``minimal" graded Anti de Sitter algebra and shows 
it to be inconsistent in its simplest form. While  one may
construct generalized algebras that still contract to  
super-Poincar\'e, these can also be shown to
fail, using for example  some results of \cite{Fre}. In \cite{sagnotti},
a Noether  procedure, starting from the full theory of 
\cite{001}, was also attempted; as we shall show below, 
there is an underlying cohomological basis for that failure.}.

We will proceed from two 
complementary starting points.  The first will
be the Noether current approach, in which we 
attempt---and fail!---to find a linearized,
``globally" supersymmetric model 
about an Anti de Sitter (AdS) background
upon which to construct a full locally
supersymmetric theory.  Since a Noether procedure
is indeed  a standard way to obtain the full 
theory, in lower dimensions,  the absence of a
starting point for it effectively  
forbids the extension. In contrast, the second
procedure will begin with the full (original)
theory of \cite{001} and attempt, using cohomology techniques, 
to construct---also unsuccessfully---a consistent
deformation of the model and of its transformation
rules that would include the desired fermion mass term 
plus cosmological term extensions.  In both cases, the
obstruction is due to the $4-$ (or $7-$) form field necessary
to balance degrees of freedom.

First, we recall some general
features relevant to the linearized approach.  
It is well-known that Einstein theory with
cosmological term linearized 
about a background solution of constant curvature  
retains its gauge 
invariance and degree of freedom count, with
the necessary modification that the vielbein
field's gauge transformation is the background
covariant  $\d h^a_\m = D_\m\xi^a$.
Similarly it is also known that the free spin
3/2 field's gauge invariance in this space
is no longer
$\d \psi_\m = \pa_\m \a (x)$ or even
$D_\m \a (x)$, but rather the extended form
\cite{003}
\be%1
\d \psi_\m = {\cal D}_\m \, \a (x) \equiv
(D_\m + m\g_\m ) \a (x)
\ee
where  ${\cal D}_\m$  has the
property that $[{\cal D}_\m , {\cal D}_\n ] =0$
when the mass $m$ is ``tuned'' to an AdS
cosmological constant: $2 m= \sqrt{-\Lambda}$
(in $D=11$).  The modified
transformation (1)
then keeps the degree of freedom count for
$\psi_\m$ the same as in flat space, 
provided---as is needed for consistency---that
the $\psi$'s action and field equations also
involve ${\cal D}_\m$ rather than $D_\m$.
[This is of course the reason for the ``mass" 
term $m \bar{\psi}_\m \G^{\m\n} \psi_\n$
acquired by the spinor field to accompany the
cosmological one for gravity.]  Given the above
facts, the 3-form potential  $A_{\m\n\rho}$
still  balances fermi/bose degrees
of freedom here.
[For now, we keep the same field content as in
the flat limit.]
Unlike the other two fields, 
its action only involves curls and so 
it neither needs nor can accomodate any extra
terms in the background to retain its gauge
invariance and excitation count; indeed, the 
only possible quadratic addition would be a
-- true -- mass term $\sim \L \, A^2$ that would
destroy both  (there would be 120, instead of the 84 
massless, excitations).  One can
therefore expect, with reason,
that the problem will lie in
the form (rather than gravity) sector's
transformation rules.  In the AdS background,
the desired ``globally" 
supersymmetric free field starting point
involves the 
Killing spinor
%``most constant" spinor 
$\e (x)$,
%namely one that 
${\cal D}_\m \e (x) =0$,
which is unrelated to the general gravitino gauge
spinor  $\a (x)$ in (1).
[Note that we can neither use $\pa_\m \e = 0$ 
because space is curved, nor
$D_\m \e =0$ because only
${\cal D}_\m$'s commute.]  The rules are 
essentially fixed from the known flat background
ones (to which they must reduce for $\L = 0$),
\begin{eqnarray}%2
\d \, \psi_\m & = & \d_h \psi_\m  + 
\d_A \psi_\m = 
\left( \frac{1}{4} X_{\m ab} (h) \G^{ab} -
m \g^a h_{\m a} \right) \e 
 +  i/144 \:
(\G^{\a\b\g\d}~_\m - 8 \, \G^{\b\g\d}
\d^\a_\m ) \e \, F_{\a\b\g\d} \; \nonumber\\
\d \, h_{\m a} & = & -i \,  \, \bar{\e} \,
\G_a\psi_\m \hspace{.4in} 
\d \, A_{\m\n\rho} = 3/2 \, \bar{\e} \,
\G_{[\m\n} \psi_{\rho ]}.
\end{eqnarray}
The linearized connection $X(h)$ is 
derived by a linearized ``vanishing torsion"
condition $D_\m h_{\n a} + X_{\m ab}
e_\n^b - (\n\m ) = 0$; throughout, the
background vielbein is $e_{\m a}$ and its
connection is $\o_{\m ab} (e)$.  Now vary the
spinorial action $I[\psi]=-1/2\int (dx) \psi_\m
\G^{\m\a\b} {\cal D}_\a \psi_\b$ (world
$\G$ indices are totally antisymmetric and 
$\G^\m = e^\m~\!_a \g^a$ etc.).  It is
easily checked that although $[\G , {\cal D}]
\neq 0$, varying $\bar{\psi}$ and $\psi$
does yield the same contribution, and using
(2) we find
\begin{eqnarray}%3
\lefteqn{\d I[\psi ] = \d_h I[\psi ] + \d_A
I [\psi ] =} \nonumber \\
& - & i/8 \int (dx) E^{\m b} (-i\k \bar{\e} \G_a 
\psi_\m ) - i/8 \int (dx) 
[ D_\a F^{\a\m\rho\s} (\bar{\e} 
\G_{[\m\n} \psi_{\rho ]} )
+ m \bar{\psi}_\m (\Gamma^{\mu\alpha\beta\rho\sigma}F_{\alpha\beta
\rho\sigma} ) \e ] \; .
\end{eqnarray}
Here $E^{\m b} $ is the variation of the 
Einstein cosmological action linearized
about AdS.  
The form-dependent piece of (3) has a first
part that behaves similarly, namely it is 
proportional to the  form field action's variation
$D_\a F^{\a\mu\rho \s}$ (the Chern--Simons
term, being cubic, is absent at this level).
With the transformation choice (2),
the variation of the Einstein plus form actions
almost  cancels (3).  There 
remains $\bar \psi F\epsilon$, the $A-$variation of
the gravitino mass term.  What possible
deformations of the transformation rules (2) and of
the actions might cancel this unwanted term?  The
only dimensionally allowed change in (2) is a
term $ \bar{\d} \psi_\m \sim m A\!\!\!/_\m
\e$; however, it will give rise to unwanted 
gauge-variant contributions from the
$m \bar{\psi} \G\psi$ term $\sim m^2 \bar{\psi}
\G A\e$, that would in turn require a true mass
term $I_m{[A]} \sim m^2 \int (dx) A^2$ to cancel,
thereby altering the degree of freedom count.
  Indeed these two deformations,
$\bar{\d} \psi_\m$ and $I_m [A]$, are the only
ones that have nonsingular
$m\rightarrow 0$ limits.  A detailed calculation 
reveals, however, that even with these added
terms, the action's invariance cannot be preserved.
In particular, there are already variations of
the $A^2$ term that cannot be compensated. 
A completely parallel calculation starting
with a dual, 7-form, model yields precisely the
same obstruction\footnote{The $7-$ form  variant was 
originally considered by \cite{Nicolai}, who
argued that it was excluded in the non-cosmological
case, but the possibility for a cosmological extension was not
entirely removed; the latter was considered 
and~~rejected~~at~~the~~Noether~~level~~in~~\cite{sagnotti}.}: 
defining the $4-$form
dual of the $7-$form, we have the same structure
as  the  $4-$form case, up to normalizations, and
face the same non-cancellation problem; also here
a mass term is useless. 

Our second approach analyses the extension problem
in the light of the
master equation and its consistent deformations 
\cite{BH,Julia,Stasheff}; see \cite{HT} for a review
of the master equation formalism appropriate to the
subsequent cohomological considerations. 
One starts with the solution of the master equation
$(S,S) = 0$ 
\cite{HT,ZJ} for the action of an undeformed theory 
(for us that of \cite{001}).
One then tries to perturb it,
$
S \rightarrow S'=S + g \Delta S^{(1)} + g^2 \Delta S^{(2)} +
.....$,
where $g$ is the deformation parameter, in such a way 
that the
deformed $S'$ still fulfills the master equation
$
(S', S') = 0. \label{masterdefo}
$
As explained in \cite{BH} any deformation of the
action of a gauge theory and of its gauge symmetries, consistent in
the sense that the new gauge transformations are indeed
gauge symmetries of the new action, leads to a deformed
solution $S'$ of the master equation. Conversely, any
deformation $S'$ of the original solution $S$ of the master
equation defines a consistent deformation of the  
original gauge invariant action and of its gauge symmetries.  
In particular, the antifield--independent  term in
$S'$ is the new, gauge-invariant action; the terms
linear in the antifields conjugate to the classical fields
define the new gauge transformations \cite{BH,GW} while the
other terms in $S'$ contain information about the deformation of
the gauge algebra and of the higher-order structure functions.
To first order in 
$g$, $(S^\prime , S^\prime)=0$ implies  $(S,\Delta S^{(1)})=0$,
{\it i.e.}, that $\Delta S^{(1)}$  (which
has ghost number zero) should be an observable 
of the undeformed
theory or equivalently
$\Delta S^{(1)}$ is ``BRST-invariant" - recall that the solution
$S$ of the master equation generates the BRST transformation in the
antibracket.  To
second order in $g$, then, we have
$
(\Delta S^{(1)},\Delta S^{(1)}) + 2
(S, \Delta S^{(2)}) = 0,  \label{obstru}
$
so the antibracket of $\Delta S^{(1)}$ with itself should be
the BRST variation of some $\Delta S^{(2)}$.  

Let us start with the full nonlinear 4-dimensional 
$N=1$ case, where a cosmological term {\it can} be
added, for contrast with $D=11$. The action is \cite{DZ1}
\be%8
I_4[e^a_\mu, \psi_\lambda] =-\frac{1}{2} \int (dx)\big(
\frac{1}{2} e e^{a \mu} e^{b \nu} R_{\mu \nu a b}
+
\overline{\psi}_\mu \Gamma^{\mu\sigma\nu} D_\sigma \psi_\nu \big),
\label{action40}
\ee
where $e \equiv \det(e_{a \mu})$ and $D_\mu$ here is of course
with respect to the full vierbein; it
is invariant  under the local supersymmetry (as well as diffeomorphism
and local Lorentz)
transformations%
%\footnote{We use first order formalism 
%in $D+4$ and second order in $D=11$, for historical
%reasons only.}
\be%9
\delta e^a_\mu =  - i\bar{\e} \Gamma^a \psi_\m \: ,
\; \;
\delta \psi_\lambda = D_\l \e (x) \: ,
\label{var1}
\ee
and under those of the spin connection $\omega^{ab}_\mu$. 
%\; \;
%\delta_\alpha \omega_{a}^{\nu \rho} = e^{-1} 
%\epsilon^{\lambda \mu \nu \rho} \bar{\alpha}
%\gamma_5 \gamma_a D_\lambda \psi_\mu.
%\label{var1}
%\ee
The solution of the master equation takes the standard form 
\be
S =  I_4 + \int\int (dx)(dy)\varphi^*_i(x) R^i_A (x,y) C^A (y)
 + X,
\label{solmaster1}
\ee
where the $\varphi^*_i$ stand for all the antifields
of antighost number one conjugate to the original
( antighost number zero) fields  $e_{a \mu}$,
$\psi_\lambda$,  and where the
$C^A$ stand for all the ghosts.  The $R^i_A (x,y)$
are the coefficients of all the gauge transformations
leaving $I_4$ invariant.  The terms
denoted by $X$ are at least of antighost number two, i.e.
contain at least two antifields $\varphi^*_i$ or one of the antifields
$C^*_\alpha$ conjugate to the ghosts.  The quadratic terms in 
$\varphi^*_i$ are also quadratic in the ghosts and arise because
the gauge transformations do not close off-shell \cite{Kallosh}.
We next recall some
cohomological background \cite{HT} related to the general solution
of the ``cocycle" condition 
$
(S,A) \equiv sA = 0
\label{cocycle}
$
for $A$ with zero ghost number.  If one expands $A$ in
antighost number
$
A = A_0 +  \bar A,
\label{expansion}
$
where $\bar A$ denotes antifield-dependent terms,
one finds that the antifield-independent term $A_0$ should be
on-shell gauge-invariant.  Conversely, given an on-shell
invariant function(al) $A_0$ of the fields, there is 
a unique, up  to irrelevant ambiguity, solution $A$
(the ``BRST invariant extension" of $A_0$)
that starts with $A_0$.
Below we shall obtain the required $A_0$.
The relevant property that makes the introduction of a cosmological
term possible in four dimensions is the fact that a gravitino mass term
$ m \int (dx) e \overline{\psi}_\lambda
\Gamma^{\lambda \rho} \psi_\rho \,  $
defines an observable; one easily
verifies that it is on-shell gauge invariant under 
(\ref{var1}).  Hence,
one may complete it with antifield-dependent terms,
to define the initial deformation $m\Delta S^{(1)}$
that satisfies $(\Delta S^{(1)},S) = 0$.
The antifield-dependent contributions are fixed by the
coefficients of the field equations in the gauge variation
of the mass term.  Specifically, since
one must use the {\it undeformed} equations
for the gravitino and the spin connection in order to verify
the invariance of the mass term under supersymmetry
transformations, these contributions will be of the form 
$\psi^* C$ and $\omega^* C$, where
$C$ is the commuting supersymmetry ghost.  They then  lead to
the known \cite{Townsend} modification of the supersymmetry
transformation rules
for the gravitino and the spin connection when the
mass term is turned on\footnote{  A complete
investigation of the BRST cohomology of $N=1$ supergravity 
has been recently
carried out in \cite{Brandt}.}.
Having obtained an acceptable first order deformation,
$m \Delta S^{(1)}$, we must in principle proceed to verify
that $(\Delta S^{(1)}, \Delta S^{(1)})$ is the BRST variation of
some $\Delta S^{(2)}$; indeed it is , with $\Delta S^{(2)}=
3/2 \int (dx) e $, as expected.
There are no higher order terms in the deformation
parameter $m$ because
the antibracket of $\Delta S^{(1)}$ with $\Delta 
S^{(2)} $ vanishes ($\Delta S^{(1)}$ does not contain 
the antifields conjugate to the vierbeins), so the 
complete solution of the master equation with 
cosmological constant is 
$
S + m \Delta S^{(1)} + m^2 \Delta S^{(2)}
$, the action of \cite{Townsend}.
[The possibility of introducing the gravitino mass
term as an observable deformation hinged on the
availability of a dynamical curved geometry in the
sense that while
$(S,\Delta S^{(1)})=0$ is always satisfied, only then is  
$(\Delta S^{(1)}, \Delta S^{(1)})$  BRST exact, i.e.
is there a second order --gravitational-- deformation.]

To summarize the analysis of the four-dimensional 
case, we stress that the cosmological term appears,
in the formulation without auxiliary fields followed here,
as the second order term of a consistent deformation of 
the ordinary supergravity action whose first order 
term is the gravitino mass term, with the
mass as deformation parameter;  
it is completely fixed by the
requirement that the deformation preserve
the master equation and
hence gauge invariance. This means, in particular, that the 
cosmological constant 
itself must be fine-tuned
to the value $-4 m^2$, as  explained 
in \cite{003}.\footnote{We emphasize that in
this procedure, one cannot 
start with the cosmological
term as a $\Delta S^{(1)}$.  Indeed, the variation 
of the cosmological term under the gauge 
transformations of the undeformed theory
is algebraic in the fields and hence does not vanish
on-shell, even up to a surface term.  Hence it is not an
observable of the undeformed theory, and so cannot be a
starting point for a consistent deformation:
adding the
cosmological term
(or the sum of it and the mass term) as a $\Delta S^{(1)}$
to the ordinary supergravity action 
is a much more radical (indeed inconsistent !) change
than the gravitino mass term alone.} 
%The cosmological term can only arise as second 
%order term, which explains, as we mentioned above, the
%fact that it is fine-tuned to the value $-m^2$ 
%in terms of the deformation parameter $m$.}

Let us now turn to the action $I_{CJS}$ of \cite{001} in
$D=11$.
The solution of the master equation again takes the standard
form\footnote{Many of the features  of (\ref{solmaster2}) were 
anticipated in \cite{deWit}.}
\be
S =  I_{CJS} 
+ \int \int (dx)(dy)\varphi^*_i(x) R^i_A (x,y) C^A (y)
+ \int (dx)C^{*\mu \nu} \partial_\mu \eta_\nu 
+ \int (dx)\eta^{* \mu} \partial_\mu \rho   + Z,
\label{solmaster2}
\ee
where the $\eta_\nu$ and $\rho$ are the ghosts of ghosts and ghost 
of ghost of ghost necessary to account for the gauge symmetries of
the $3$-form $A_{\lambda \mu \nu}$, and where $Z$ (like $X$
in (6)) is  determined
from the  terms written by the  $(S,S) = 0$ requirement.
As in $D=4$, we seek a first-order deformation analogous to 
\be
 \Delta S^{(1)} =   \frac{1}{2} m\int (dx) e \overline{\psi}_\lambda
\Gamma^{\lambda \rho}
\psi_\rho \,   +
\hbox{ antifield-dep.} \label{mass2}
\ee
However, contrary to what happened at $D=4$, the
mass term no longer defines an observable, as
its variation under local supersymmetry
transformations reads
\be
\delta \big( e \overline{\psi}_\lambda
\Gamma^{\lambda \rho}
\psi_\rho \big) 
\approx  -\frac{i}{18}\overline{\psi}_\mu \Gamma^{\mu
\alpha \beta \gamma \delta} \e F_{\alpha \beta \gamma \delta}
+ O(\psi^3) 
\label{fail}
\ee
where $\approx$ means equal on shell up to a divergence.
%The right-hand side of
%(\ref{fail}) does not vanish on-shell, even up to a total 
%time derivative.
Indeed, the condition that the r.h.s. of (\ref{fail})
 also weakly vanish 
is easily verified to imply, upon 
expansion in the
derivatives of the gauge parameter $\e$, 
that $\overline{\psi_\mu} \Gamma^{\mu\alpha 
\beta \gamma \delta}
\e F_{\alpha \beta \gamma \delta}$ must
vanish on shell, which it does {\it not} do.

Can one improve the first-order deformation 
(\ref{mass2}) to make it acceptable?  The 
cosmological term
will not help because it does not transform 
into $F$.  The
only possible candidates would be functions 
of the 3-form
field.  In order to define observables, these functions must
be invariant under the gauge transformations of the
3-form, at least on-shell and up to a total
derivative.  However, in 11 dimensions, the only such
functions can be redefined so as to be
off-shell (and not just on-shell) gauge invariant,
up to a total derivative.  This follows from an argument that
closely patterns the
analysis of \cite{pform}, defining the very restricted 
class of on-shell invariant vertices  that cannot  in
general be extended  off-shell. [The above result actually justifies 
the usual assumptions, {\it e.g.}, those of 
\cite{sagnotti} that ``on--'' implies ``off--''.]
Thus, the
available functions of $A$ may be assumed to be
strictly gauge invariant, i.e., to be functions
of the field strength  $F$ (which
eliminates $A^2$; also,  changing the coefficient
of the  Chern-Simons
term in the original action clearly cannot  help).  
But it is easy to see that no
expression in $F$ can cancel the unwanted term in
(\ref{fail}),
because of a mismatch in the number of derivatives.
Hence, there is no way to improve the mass term to turn 
it into an observable
in 11 dimensions.  It is the $A$-field part of the
supersymmetry variation of the gravitino that
is responsible for the failure of the mass term to
be an observable, just as it was also responsible for the
difficulties described in the first, linearized, approach. 
Since the cohomology procedure saves us from also seeking modifications
of the transformations rules,
we can conclude that the 
introduction of a cosmological constant is 
obstructed already at the first
step in $D=11$ supergravity from the full theory
end as well.

In our discussion, we have assumed (as in
lower dimensions) both that the limit of a vanishing mass $m$ is
smooth\footnote{ This restriction is not necessarily stringent:
in cosmological $D=10$ supergravity \cite{Roman}, there
is  $m^{-1}$ dependence in a field
transformation rule, but that is an artefact removable
by  introducing a Stuckelberg compensator.}
 and that the field content remains unchanged
in the cosmological variant. Any ``no-go''  result
is of course no stronger than its assumptions, and
ours are shared by the earlier treatments
\cite{Nahm,sagnotti} that we surveyed. There is
one (modest) loosening that can be shown not to
work either, inspired by a recent reformulation
\cite{Green} of the $D=10$ cosmological model 
\cite{Roman}. The idea is to add a deformation involving
a nonpropagating field, here the 11-form $G_{11}\equiv
d A_{10}$, through an addition $\Delta I \sim
\int (dx)  [G_{11}+b \bar\psi
\Gamma^9\psi]^2$. The $A_{10}$-field equation states that
the dual, $\epsilon^{11} [ G_{11}+b
\bar\psi \Gamma^9\psi]$ is a constant of integration,
say $m$. The resulting supergravity field equations look
like the ``cosmological'' desired ones. However, while
this ``dualization'' works for lower dimensions, in $D=11$ we are simply
back to the original inconsistent model with
supersymmetry still irremediably lost, as  can be also discovered 
--without integrating out-- in the deformation approach.

\bigskip
The work of S. D. and D. S. was supported by  
NSF grant PHY 93-15811

\end{document}